\documentclass{ws-p10x7}
\global\arraycolsep=2pt

\begin{document}

\title{New Limits on the Production of Magnetic Monopoles at Fermilab}

\author{K. A. Milton, G. R. Kalbfleisch, M. G. Strauss, 
L. Gamberg,  W. Luo, and E. H. Smith}

\address{Department of Physics and Astronomy, University of Oklahoma,
Norman, OK 73019-0225 USA}



\twocolumn[\maketitle\abstract{First results from an experiment (Fermilab E882)
searching for magnetically
charged particles bound to elements from the CDF and D\O\ detectors are 
reported.  The experiment is described, and limits on magnetic 
monopole pair production cross sections for magnetic charges 1, 2, 3, 
and 6 times the Dirac pole strength are presented. These limits ($\sim1$ pb),
hundreds of times smaller than those found in previous direct
accelerator-based searches,
use simple model assumptions for the photonic production of monopoles, as
does the extraction of mass limits in the hundreds 
of GeV range.}]

\section{Introduction}


The most obvious reason for introducing magnetic charge 
into electrodynamic theory is the symmetry thereby
imparted to Maxwell's equations.  
Further, the introduction of fictitious magnetic charge 
simplifies many calculations,
as Bethe and Schwinger realized in their work on waveguides during World War 
II.\cite{bethejs}


Henri Poincar\'e first studied the classical dynamics of an electron
moving in the field of a magnetic monopole,\cite{poincare} while J. J.
Thomson in lectures at Yale demonstrated that a classical 
static system consisting of electric ($e$) and magnetic ($g$) charges separated
by a distance $R$ had an intrinsic angular momentum pointing along the line
separating the charges\cite{thomson}:
${\bf J}
={eg\over c}{\bf \hat R}$.
Requiring that the radial component of this angular momentum be a multiple 
of $\hbar/2$ leads to Dirac's celebrated quantization condition,
$eg={n\over2}\hbar c, \quad n=\pm1,\pm2,\pm3,\dots$.
In fact, Dirac obtained this quantization condition by showing
that quantum mechanics with magnetic monopoles was consistent only if
this quantization condition held.\cite{dirac}  Thus, the existence of a
single monopole in the universe would explain the empirical fact of
the quantization of electric charge.  Schwinger generalized this
quantization condition to dyons, particles carrying both electric and
magnetic charge.\cite{schwingerdyon} 
He further argued that $n$ had to be an even integer (sometimes even 4 times
an integer).\cite{schwingerdyon} Thus the smallest positive
value of $n$ could be 1 or 2, or 3 or 6 if it is the quark electric charge which
quantizes magnetic charge.

\section{Experiment Fermilab E882}
The concept of the present experiment is that low-mass
monopole--anti-monopole pairs could be produced by the proton--anti-proton
collisions at the Tevatron.  The monopoles produced would travel only a
short distance through the elements of the detector surrounding the
interaction vertex before they would lose their kinetic energy and become
bound to the magnetic moments of the nuclei in the material making up
the detector.  We have obtained a large portion of the old detector elements
(Be, Al, Pb) from the D\O\ and CDF experiments, and are in the process
of searching for monopoles in these materials using an induction detector.
A first paper describing our analysis of a large part of the D\O\ Al and Be
samples has appeared.\cite{E882}

The model for the production process is that the monopole pairs are produced
through a Drell-Yan process, which includes one factor of the velocity $\beta$
to account for the phase space, and two additional factors of $\beta$ to
simulate the velocity suppression of the magnetic coupling.  We
use this rather simple model, the best available,
because a proper field theoretical description
of monopole interactions still does not exist.\cite{gamberg}

Any monopoles produced by the Tevatron are trapped in surrounding detector
elements with 100\% probability, and will be bound in that material
permanently provided it is not melted down or dissolved.\cite{fest}
Although the theory of binding is also in a crude state, monopole binding
energies to nuclei are at least in the keV range, which is of the same order
as the energy trapping the nucleus-monopole complex to the material lattice,
more than adequate to insure permanent binding (and to preclude the extraction
of monopoles from the sample by available magnetic fields).
 
We can set much better limits than those given by previous 
direct accelerator-based searches\cite{bertani} 
because the integrated luminosity of Fermilab has increased by
a factor of about $10^4$ to $172\pm8$ pb$^{-1}$ for D\O.

A schematic of the apparatus is available: {\it
 www.nhn.ou.edu/\%7Egrk/apparatus.pdf}.  The Fermilab samples are cut
to a size of approximately $(7.5 \mbox{ cm})^3$ and are repeatedly moved up and
 down through a warm bore in a magnetically shielded
 cryogenic detector.  The active elements are
 two superconducting loops connected to SQUIDs, which convert any current
 in the loops into a voltage signal.  In empty space, the persistent 
 current set up in the loop having inductance $L$ by a monopole of charge $g$
 passing through it is 
$ LI=4\pi g/ c$.
A more exact expression was used in fitting data.

\begin{figure}
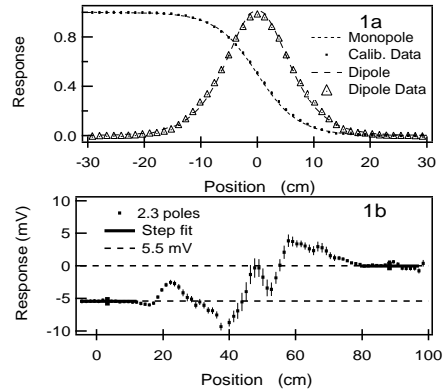

\epsfxsize15pc
\figurebox{16pc}{16pc}{Layout0_Fig1AB.EPSF}
\caption{``Pseudopole'' curves.  
a) Comparison of theoretical monopole response  to an 
experimental calibration and of
 a simple point dipole of one sample with that calculated 
from the theoretical response curve.  b) The observed 
``step'' for a pseudopole current, corresponding to 2.3 minimum 
Dirac poles, embedded in an Al sample.}
\label{fig1}
\end{figure}

A pseudopole was constructed by making a long solenoid, which could either
be physically moved through the detector loop, or turned on and off.  It was
also attached to an actual sample, so the background due to magnetic
dipoles in the sample could
be seen.  Results of such tests are shown in Fig.~\ref{fig1}.  This demonstrates that
we could easily detect a Dirac monopole.

\begin{figure}
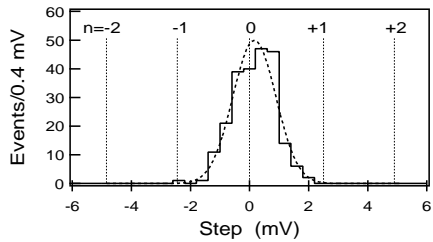

\epsfxsize15pc
\figurebox{16pc}{32pc}{Layout0_Fig3.EPSF}
\caption{Histogram of steps.  Vertical lines define the expected positions
of signals for various $n$.}
\label{fig3}
\end{figure}

The monopole signal is a step in the output of the SQUID after that
output has returned from its relatively large excursions resulting from dipoles
in the sample.  222 Al and 6 Be samples were analyzed, and the distribution
of steps had a mean of 0.16 mV and and rms spread of 0.73 mV, as shown in
Fig.~\ref{fig3}.  We use the Feldman-Cousins analysis\cite{fc}: Because
8 samples were found within 1.28 $\sigma$ of $n=\pm1$, where 10.4 were
expected, we can say at the 90\% confidence level that the upper limit
to the number of signal events with $n=\pm1$ is 4.2.  We also remeasured
those outlying events, and found that all were within $2\sigma$ of $n=0$,
so we have no monopole candidates in this set.  Similarly, the upper limit
to the number of $|n|\ge2$ events is 2.4.

\begin{table*}[t]
\caption{Acceptances, upper cross section limits, and lower mass limits,
as determined in this work (at 90\% CL).}
\begin{tabular}{|ccccc|}
\hline
Magnetic Charge&$|n|=1$&$|n|=2$&$|n|=3$&$|n|=6$\\
\hline
Sample&	Al&Al&Be&Be\\
$\Delta\Omega/4\pi$ acceptance&0.12&0.12&0.95&0.95\\
Mass Acceptance&0.23&0.28&0.0065&0.13\\
Number of Poles&$<4.2$&$<2.4$&$<2.4$&$<2.4$\\
Upper limit on cross section&0.88 pb&0.42 pb&2.3 pb&0.11 pb\\
Monopole Mass Limit&$>285$ GeV&$>355$ GeV&$>325$ GeV&$>420$ GeV\\
\hline
\end{tabular}
\label{tab1}
\end{table*}

We use the $\beta^3$ modified Drell-Yan production model 
together with the evolved CTEQ5m parton distribution functions\cite{cteq} 
to estimate the acceptance of our experiment, as shown in  Table~\ref{tab1}.  
Using the total luminosity delivered to D\O,
the number limit of monopoles, the mass acceptance so calculated, and the
solid angle coverage of our samples, we can obtain the $p\bar p$ cross
section limits for the production of monopoles as given in Table~\ref{tab1}.  
These
are better by a factor of 200 than the earlier results of Bertani.\cite{bertani}
Using the production model again, we can convert these cross section limits
into mass limits, simply by scaling the Drell-Yan cross section by the
monopole enhancement factor of $n^2(137/2)^2$. These mass limits,
also indicated in Table~\ref{tab1}, are some  3 times 
larger than those of prior searches for accelerator-produced
monopoles trapped in matter.\cite{bertani}

\section{Conclusions}
This experiment to detect low-mass, accelerator produced magnetic monopoles
is continuing.  Over the next year, we will analyze the remaining Pb and
Al samples from CDF, and extract better limits.  In the future, such an
experiment carried out using LHC exposures could reach monopole masses of
a few TeV.

\section*{Acknowledgments}
This work was supported in part by the US Department of Energy.

\end{document}